\documentclass[journal]{IEEEtran}
\usepackage[T1]{fontenc}
\usepackage{amsmath}
\usepackage{amsfonts} 
\usepackage{color}
\usepackage{pgfplots}
\usepackage{cite}
\pgfplotsset{compat=1.5}
\usepackage{pgfplotstable}
\usepgfplotslibrary{statistics}
\pgfplotsset{grid style={dotted,gray}}
\usepackage{tikz}

\begin{document}
\title{Autoencoder-Based Unequal Error Protection Codes}

\author{Vukan Ninkovic, ~\IEEEmembership{Student Member,~IEEE,} Dejan Vukobratovic, ~\IEEEmembership{Senior Member,~IEEE,} Christian H\"{a}ger, ~\IEEEmembership{Member,~IEEE,} Henk Wymeersch, ~\IEEEmembership{Senior Member,~IEEE,} Alexandre Graell i Amat, ~\IEEEmembership{Senior Member,~IEEE} 

\thanks{V. Ninkovic and D. Vukobratovic are with the Department of Power, Electronics and Communications Engineering, University of Novi Sad, 21000, Novi Sad, Serbia (e-mail: \{ninkovic, dejanv\}@uns.ac.rs).}

\thanks{C. H\"{a}ger, H. Wymeersch, and A. Graell i Amat are with the Department of Electrical Engineering, Chalmers University of Technology, SE–41296 Gothenburg, Sweden (e-mail: \{christian.haeger, henkw, alexandre.graell\}@chalmers.se).}

\thanks{This work has received funding from the European Union's Horizon 2020 research and innovation programme under Grant Agreement number 856967.}
}

\maketitle

\begin{abstract}
We present a novel autoencoder-based approach for designing codes that provide unequal error protection (UEP) capabilities. The proposed design is based on a generalization of an autoencoder loss function that accommodates both message-wise and bit-wise UEP scenarios. In both scenarios, the generalized loss function can be adjusted using an associated weight vector  to trade off error probabilities corresponding to different importance classes. For message-wise UEP, we compare the proposed autoencoder-based UEP codes with a union of random coset codes. For bit-wise UEP, the proposed codes are compared with  UEP rateless spinal codes and the superposition of random Gaussian codes. In all cases, the autoencoder-based codes show superior performance while providing design simplicity and flexibility in trading off error protection among different importance classes. 
\end{abstract}


\IEEEpeerreviewmaketitle

\section{Introduction}
\label{intro}

Learning transmitters and receivers for a given channel model using deep autoencoders (AE) optimized for a specific loss function, e.g., the one targeting minimized block-error rate (BLER), is recently proposed in \cite{OShea_2017}. Such AE-based encoders and decoders achieve close-to-optimal performance for some baseline communication scenarios \cite{OShea_2017, OShea_2017_2}, and have been demonstrated in a proof-of-concept real-world implementation \cite{Dorner_2018}. The works \cite{OShea_2017, OShea_2017_2, Dorner_2018} and other recent works consider AE-based encoders and decoders that provide equal error protection across the set of transmitted messages.  However, in many communication scenarios such as transmission of control signals along data, multi-resolution source coding, and ultra-reliable and low-latency communication protocols, one is interested in the design of codes that provide unequal error protection (UEP) capabilities \cite{Masnick_1967}.

UEP codes are commonly investigated in two different scenarios: message-wise UEP and bit-wise UEP \cite{Shkel_2015,Borade_2009}. In message-wise UEP, the set of source messages is divided into disjoint subsets or importance classes, each of which may be provided with a different level of error protection. Although certain asymptotic and non-asymptotic results on message-wise UEP are derived in \cite{Shkel_2015}, explicit constructions of message-wise UEP codes are rare. In bit-wise UEP, a message is encoded into a sequence of bits and different subblocks of bits represent different importance classes that are protected differently \cite{Borade_2009}. Practical designs of bit-wise UEP codes are more frequent in the literature, due to their applicability in, e.g., improved packet header protection or multimedia communication protocols. For example, bit-wise UEP has been proposed for coding schemes such as LDPC \cite{Pishro-Nik_2005}, fountain \cite{Sejdinovic_2009}, or spinal codes \cite{Yu_2016}.

In this work, we propose a novel AE-based approach to design codes that yield UEP capabilities. In the proposed design, we introduce a novel AE compound loss function that consists of a weighted contribution of each importance class. The proposed approach is generic enough to allow for the design of both AE-based message-wise and bit-wise UEP codes. For both message-wise and bit-wise UEP, we compare the performance of the proposed AE-based codes with that of conventional UEP code designs in the literature. For the case of message-wise UEP, our AE-based codes significantly outperform the construction considered in  \cite{Shkel_2015} based on the union of coset codes. For bit-wise UEP, the proposed AE-based codes outperform  UEP rateless spinal codes \cite{Yu_2016} and  UEP codes based on the superposition of random Gaussian codes \cite{Karimzadeh_2019}. Moreover, for the proposed AE-based codes, a simple and flexible procedure of fine-tuning the weight parameters provides a desired trade-off among error probabilities of different importance classes. 

\section{Background}
\label{bckgnd}

\subsection{System Model}
\label{sec:Model}

We consider the problem of communicating a message $m$ from a set of messages $\mathcal{M}=\{1,2,\ldots,M\}$ over a noisy channel. Each message  is represented as a sequence of bits $\boldsymbol{s}=(s_1,s_2,\ldots,s_k)$, where $k=\log_2(M)$ is the message length. We define the encoder mapping $f:\mathcal{M} \rightarrow \mathbb{R}^n$ that encodes the message $m$ into a codeword $\boldsymbol{x}=(x_1,x_2,\ldots,x_n)$ of length $n$. The transmitted codewords obey the total energy constraint $\| \boldsymbol{x} \|_2^2=n$. The code rate is  $R=k/n$ (in bits per channel use). The channel $\mathcal{W}$ transforms the input codeword $\boldsymbol{x} \in \mathbb{R}^n$ into the output sequence $\boldsymbol{y} \in \mathbb{R}^n$ following the probabilistic channel law $p(\boldsymbol{y}|\boldsymbol{x})$. Finally, the decoder mapping $g: \mathbb{R}^n \rightarrow \mathcal{M}$ produces an estimate $\hat{m}$ of the transmitted message $m$. Under the above setup, the goal is to design a pair $(f,g)$ for the channel $\mathcal{W}$  to minimize the average message error probability
\begin{align}
P_{\textrm{e}} = \frac{1}{M} \sum_{m \in \mathcal{M}} \mathbb{P}\{\hat{m} \neq m|m\}.   
\end{align}

\begin{figure}
	\centering
	\includegraphics[width=0.8\linewidth]{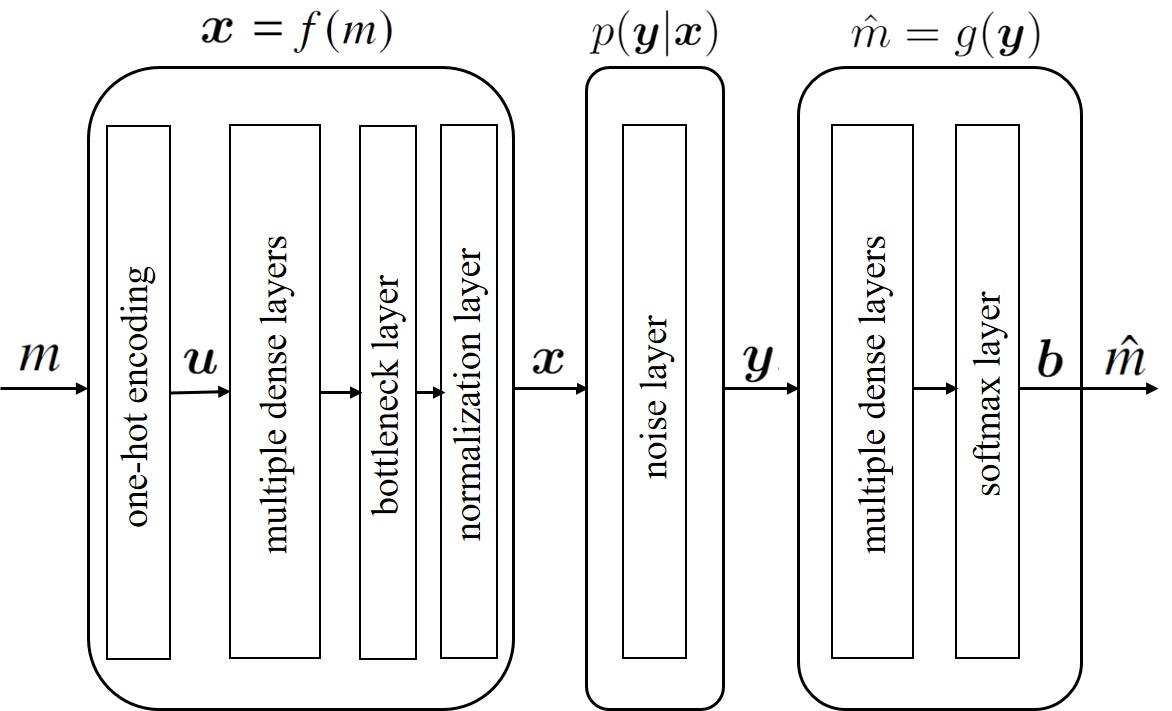}
	\caption{Communication system represented as a deep autoencoder \cite{OShea_2017}.}
	\label{Fig_AE}
\end{figure}

\subsection{Autoencoder-Based Code Design}

From a deep learning perspective, the above communication system can be represented as an AE \cite{OShea_2017}. An AE consists of a set of encoder layers representing the encoder mapping $\boldsymbol{x}=f(m)$, the noise layer modeling the channel $\mathcal{W}$ that transforms $\boldsymbol{x}$ into $\boldsymbol{y}$, and a set of decoder layers representing the decoder mapping $\hat{m}=g(\boldsymbol{y})$, as shown in Fig. \ref{Fig_AE}. 

At the input of the encoder layers, the message $m$ is encoded as a one-hot vector $\boldsymbol{u} = (u_1,u_2,\ldots,u_M) \in \{0,1\}^M$, i.e., it is represented as an $M$-dimensional vector with the $m$-th element equal to one and the others equal to zero. The set of encoder layers represents a feed-forward neural network with $H$ hidden layers, followed by a bottleneck layer of width $n$. The goal of the encoder neural network is to find the most suitable representation of the information so that it is robust to the channel perturbations. At the output of the bottleneck layer, normalization ensures that the fixed energy constraint on $\boldsymbol{x}$ is met. 

Next, the codewords $\boldsymbol{x}$ are passed through a noise layer that represents the channel $\mathcal{W}$. In this paper, we consider an additive white Gaussian noise (AWGN) channel, thus at the output of the noise layer we have $\boldsymbol{y}=\boldsymbol{x}+\boldsymbol{n}$, where $\boldsymbol{n}$ contains $n$ independent and identically distributed (i.i.d.) samples of a Gaussian random variable $\mathcal{N}(0,\sigma^2)$ of variance $\sigma^2$. 

The output of the noise layer $\boldsymbol{y}$ is fed into the decoder layers representing the receiver. The receiver is implemented in the same way as the transmitter via a feed-forward neural network, except that the last layer has a softmax activation function with output $\boldsymbol{b} = (b_1,b_2,\ldots,b_M) \in (0,1)^M$, where the $\ell_1$ norm $\Vert \boldsymbol{b}\Vert_1=1$. The decoded message is $\hat{m}=\arg\max_{i}\{b_i\}$. Except for the last layer at the transmitter and the receiver that have a linear and a softmax activation function, respectively, all others layers are activated by a rectified linear unit (ReLU).

The AE is trained in an end-to-end manner by using stochastic gradient descent (SGD) with the Adam optimizer \cite{adam} on the set of all possible messages $m \in \mathcal{M}$. The minimization of the cross-entropy loss between $\boldsymbol{u}$ and $\boldsymbol{b}$ is used as a surrogate for minimizing the error probability $P_{\textrm{e}}$, which cannot be used directly as it is not differentiable. The AE is trained using batches of training data by minimizing the cross-entropy loss function 
\begin{align}\label{eq2}
\ell(\boldsymbol{u},\boldsymbol{b})=-\sum_{i=1}^{M} u_i \log{b_i},    
\end{align}
 averaged across a batch of training samples. It is important to note that the random nature of the channel acts as a form of regularization, because the receiver never sees the same training example twice. As a consequence, it is almost impossible for the neural network to overfit \cite{Dorner_2018}.

\subsection{Unequal Error Protection Code Design}

We consider both message-wise UEP and bit-wise UEP, as detailed below. 

\subsubsection{Message-wise UEP}

We assume that the message set $\mathcal{M}$ containing $M$ messages is partitioned into $C \leq M$ disjoint subsets, referred to as message classes, having different error protection requirements. Message class $\mathcal{M}_i$ contains $|\mathcal{M}_i|=M_i$ messages, with $M=\sum_{i=1}^{C} M_i$. For a given encoder-decoder pair $(f,g)$, we define the per-class probability of error
\begin{align}
P_{\textrm{e}}^{(i)}=\frac{1}{M_i}\sum_{m \in \mathcal{M}_i} \mathbb{P}\{\hat{m} \neq m| m\}.
\end{align}
We collect the per-class error probabilities of a message-wise UEP code  in a vector  $\boldsymbol{P}_{\textrm{e}}=(P_{\textrm{e}}^{(1)}, P_{\textrm{e}}^{(2)}, \ldots, P_{\textrm{e}}^{(C)})$. We use the term message-wise UEP code to refer to the triple $(\{\mathcal{M}_i\}_{i=1}^C,f,g)$ \cite{Shkel_2015}.

Let $\mathcal{P}_{\mathcal{W}}(\{\mathcal{M}_i\}_{i=1}^C,n)\subset [0,1]^C$ denote the region of achievable $\boldsymbol{P}_{\textrm{e}}$-values for message-wise UEP codes of length $n$ and the message classes $\{\mathcal{M}_i\}_{i=1}^C$ over the channel $\mathcal{W}$. We note that the characterization of $\mathcal{P}_{\mathcal{W}}(\{\mathcal{M}_i\}_{i=1}^C,n)$ for message-wise UEP using  tools from finite-length information theory is considered in \cite{Shkel_2015}, albeit only for the binary symmetric channel (BSC) and the binary erasure channel (BEC).

\subsubsection{Bit-wise UEP} 

In bit-wise UEP, we consider an equivalent representation $\mathcal{S}$ of the message set $\mathcal{M}$ that consists of the binary representation ($\boldsymbol{s}$, see Section~\ref{sec:Model}) of the messages $m \in \mathcal{M}$. Further, we assume that $\boldsymbol{s}$ consists of $C$ submessages representing disjoint subsequences of bits, i.e., $\boldsymbol{s}=(\boldsymbol{s}_1,\boldsymbol{s}_2,\ldots,\boldsymbol{s}_C)$, where the length of submessage $\boldsymbol{s}_i$ is $k_i$ bits and $k = \sum_{i=1}^C k_i$. We denote by $\mathcal{S}$ (respectively, $\mathcal{S}_i$) the set of all possible binary messages $\boldsymbol{s}$ (submessages $\boldsymbol{s}_i$), with $|\mathcal{S}| = 2^{k}$ ($|\mathcal{S}_i| = 2^{k_i}$). In the case of bit-wise UEP, the different submessages represent the different message classes and are assigned different error protection requirements. 

We are interested in the probability of error associated with a particular message class. For a submessage $\boldsymbol{s}_i \in \mathcal{S}_i$, we denote by $\mathcal{M}_{\boldsymbol{s}_i}$ the set of all messages $m \in \mathcal{M}$ whose $i$-th submessage in the corresponding representation $\boldsymbol{s}$ equals $\boldsymbol{s}_i$. For a given encoder-decoder pair $(f,g)$, we define the per-class probability of error
\begin{align}
P_{\textrm{e}}^{(i)}=\frac{1}{|\mathcal{S}_i|} \sum_{\boldsymbol{s}_i \in \mathcal{S}_i} \mathbb{P}\{\hat{m} \notin \mathcal{M}_{\boldsymbol{s}_i}| m \in \mathcal{M}_{\boldsymbol{s}_i}\}, 
\end{align}
and define  $\boldsymbol{P}_{\textrm{e}}=(P_{\textrm{e}}^{(1)}, P_{\textrm{e}}^{(2)}, \ldots, P_{\textrm{e}}^{(C)})$. We use the term bit-wise UEP code to refer to the triple $(\{\mathcal{S}_i\}_{i=1}^C,f,g)$.

For some practical applications that call for bit-wise UEP, the above definition of per-class error probabilities does not capture the code design requirements. For example, of great interest is the case where the messages are encoded in such a way that the importance of submessages progressively decreases from $\mathcal{S}_1$ to $\mathcal{S}_C$ and there exists an interdependence between importance classes. More precisely, in such applications, the $i$-th message block $\boldsymbol{s}_i$ is considered as correctly received if and only if this block as well as all blocks $\boldsymbol{s}_j$ for $j < i$ are decoded correctly \cite{Chande_2000}. To accommodate for this scenario, we redefine the per-class error probability as
\begin{align}
P_{\textrm{e}}^{(i)}=&\frac{1}{|(\mathcal{S}_1,\ldots,\mathcal{S}_i)|} \cdot\\ \nonumber
& \sum_{\boldsymbol{s}_1,\ldots,\boldsymbol{s}_i \in (\mathcal{S}_1,\ldots,\mathcal{S}_i)} \mathbb{P}\{\hat{m} \notin \mathcal{M}_{\boldsymbol{s}_1,\ldots,\boldsymbol{s}_i}| m \in \mathcal{M}_{\boldsymbol{s}_1,\ldots,\boldsymbol{s}_i}\},    
\end{align}
where $\mathcal{M}_{\boldsymbol{s}_1,\ldots, \boldsymbol{s}_i}$ is the set of all messages $m \in \mathcal{M}$ whose binary representation $\boldsymbol{s}$ is consistent with the first $i$ submessages $\boldsymbol{s}_1,\ldots,\boldsymbol{s}_i$. We refer to this case as \emph{progressive} bit-wise UEP.

\section{Autoencoder-Based UEP Codes}
\label{uep-ae}

\subsection{UEP Autoencoders}

In this section, we present a flexible and efficient method to design encoders and decoders for both message-wise and bit-wise UEP codes by training deep AEs. The key idea of the proposed AE-based design is to define an appropriate \emph{compound loss function} that generalizes the cross-entropy loss function defined in \eqref{eq2} to the UEP case.

\subsubsection{Message-wise UEP}

Let $\ell_{\mathcal{M}_j}(\boldsymbol{u},\boldsymbol{b})$ be the loss function associated to the $j$-th message class defined as
\begin{align}
\ell_{\mathcal{M}_j}(\boldsymbol{u},\boldsymbol{b}) = -\sum_{i \in \mathcal{M}_j} u_i \log{b_i}.    
\end{align}
We redefine the loss function for the case of message-wise UEP as the weighted sum of the loss functions
$\ell_{\mathcal{M}_j}(\boldsymbol{u},\boldsymbol{b})$ as
\begin{align}
    \ell(\boldsymbol{u},\boldsymbol{b})=\sum_{j=1}^C \lambda_j \ell_{\mathcal{M}_j}(\boldsymbol{u},\boldsymbol{b}),
\end{align}
where $\pmb{\lambda} = (\lambda_1, \lambda_2, \ldots, \lambda_C)$ is a weight vector associated to the message classes, $\sum_{j=1}^C \lambda_j = 1$, and $\lambda_j \geq 0$. The rest of the training procedure follows the same steps as for the standard (equal error protection) AE-based codes.

\subsubsection{Bit-wise UEP}

We use the same approach of modifying the loss function using a weighting vector. In this case, however, some updates in the message representation are needed beforehand. We extend the definition of one-hot vector $\boldsymbol{u}$ so that it indicates a subset of messages in $\mathcal{M}$ whose binary symbol representation $\boldsymbol{s}$ is consistent with a given submessage $\boldsymbol{s}_j \in \mathcal{S}_j$. More precisely, for every submessage $\boldsymbol{s}_j \in \mathcal{S}_j$, we define the corresponding vector $\boldsymbol{u}_{\boldsymbol{s}_j} = (u_1,u_2,\ldots,u_M)$, with a $1$ in the $m$-th position if the message $m$ is such that the $j$-th submessage of its binary sequence representation is $\boldsymbol{s}_j$. Note that $\boldsymbol{u}_{\boldsymbol{s}_j}$ is now a binary vector with $2^{k-k_j}$ ones. Let $\ell(\boldsymbol{u}_{\boldsymbol{s}_j},\boldsymbol{b})$ be the loss function associated to the $j$-th submessage,
\begin{align}\label{eq4}
\ell(\boldsymbol{u}_{\boldsymbol{s}_j},\boldsymbol{b}) = -\sum_{i=1}^M u_i \log{b_i}. 
\end{align}
Given the binary sequence representation $\boldsymbol{s} = (\boldsymbol{s}_1, \boldsymbol{s}_2, \ldots, \boldsymbol{s}_C)$ of a message $m \in \mathcal{M}$, we define a set of $C$ vectors $\mathcal{U} = \{\boldsymbol{u}_{\boldsymbol{s}_1}, \boldsymbol{u}_{\boldsymbol{s}_2}, \ldots, \boldsymbol{u}_{\boldsymbol{s}_C}\}$. With the above definition, the compound loss function for bit-wise UEP is defined as
\begin{align}\label{eq3}
    \ell(\mathcal{U},\boldsymbol{b})=\sum_{j=1}^C \lambda_j \ell(\boldsymbol{u}_{\boldsymbol{s}_j},\boldsymbol{b}),
\end{align}
where  $\pmb{\lambda} = (\lambda_1, \lambda_2, \ldots, \lambda_C)$ is a weight vector associated to message classes, $\lambda_j>0$, and $\sum_{j=1}^C \lambda_j = 1$. The rest of the training procedure follows the same steps as for the standard (equal error protection) AE-based codes. 

\begin{figure}[t]
	\begin{tikzpicture}
  	\begin{loglogaxis}[width=0.95\columnwidth, height=8.3cm, 
	legend style={at={(axis cs: 0.0000012,0.0000012)}, anchor= north east,font=\scriptsize,},
   	legend cell align={left},
   	x tick label style={/pgf/number format/.cd,fixed,
   	 precision=1, /tikz/.cd},
   	y tick label style={/pgf/number format/.cd,fixed, precision=1, /tikz/.cd},
   	xlabel={$P_{\textrm{e}}^{(1)}$},
   	ylabel={$P_{\textrm{e}}^{(2)}$},
   	label style={font=\footnotesize},
   	grid=major,   	
   	xmin =0.000001, xmax = 1,
   	x dir=reverse,
   	ymin=0.000001, ymax=1,
   	y dir=reverse,
   	line width=0.8pt,
   	tick label style={font=\footnotesize},]
    \addplot[blue, mark=o] 
   	table [x={x}, y={y}] {./Fig/Fig. 1/1db.txt}; 
   	\addlegendentry{$E_\mathrm{b}/N_0$ = 1 dB} 
    \addplot[forget plot, dashed,blue, mark=o, mark options={solid}]
   	table [x={x}, y={y}] {./Fig/Fig. 2/1db.txt}; 
   	\addplot[red, mark=x] 
   	table [x={x}, y={y}] {./Fig/Fig. 1/3db.txt}; 
   	\addlegendentry{$E_\mathrm{b}/N_0$ = 3 dB} 
    \addplot[forget plot, dashed,red, mark=x, mark options={solid}]
   	table [x={x}, y={y}] {./Fig/Fig. 2/3db.txt}; 
   	\addplot[green, mark=square] 
   	table [x={x}, y={y}] {./Fig/Fig. 1/5db.txt}; 
   	\addlegendentry{$E_\mathrm{b}/N_0$ = 5 dB} 
    \addplot[forget plot, dashed,green, mark=square, mark options={solid}]
   	table [x={x}, y={y}] {./Fig/Fig. 2/5db.txt}; 
   	\addplot[purple, mark=triangle] 
   	table [x={x}, y={y}] {./Fig/Fig. 1/7db.txt}; 
   	\addlegendentry{$E_\mathrm{b}/N_0$ = 7 dB} 
    \addplot[forget plot, dashed,purple, mark=triangle, mark options={solid}]
   	table [x={x}, y={y}] {./Fig/Fig. 2/7db.txt}; 
   	\node at (-3.5,-11.9) {\small $\lambda=0.1$};
   	\node at (-8.8,-8.4) {\small $\lambda=0.5$};
   	\node at (-11,-2.7) {\small $\lambda=0.9$};
  	\end{loglogaxis}
	\end{tikzpicture}
	\vspace*{-8mm}
	\caption{$(P_\mathrm{e}^{(1)},P_\mathrm{e}^{(2)})$ performance of AE-based message-wise (solid curves) and bit-wise UEP codes (dashed curves) with $C=2$, $|\mathcal{M}_1|=8, |\mathcal{M}_2|=8$, $n=7$, for $E_\mathrm{b}/N_0 = \{1,3,5,7\}$~dB and $\lambda=\{0.1,0.2,\ldots,0.9\}$.}
	\label{Fig_2}
\end{figure}
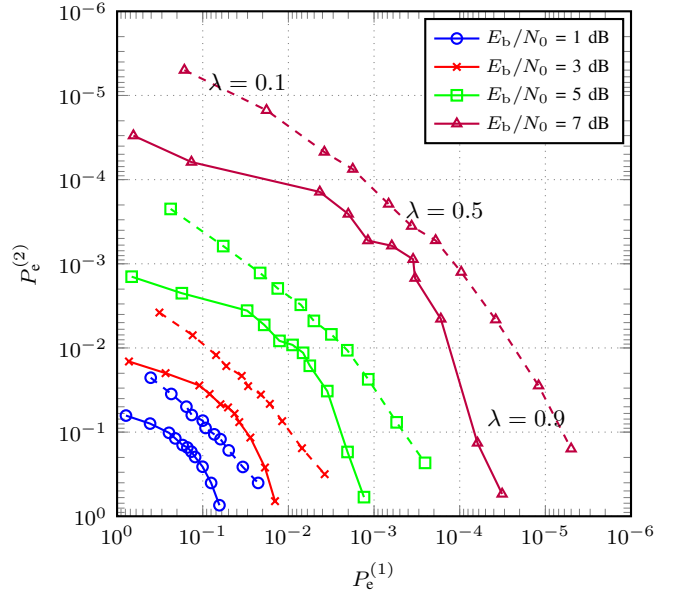

The case of progressive bit-wise UEP can be treated in a similar way. For every submessage $\boldsymbol{s}_i \in \mathcal{S}_i$, we define the binary vector $\boldsymbol{u}_{\boldsymbol{s}_1,\ldots,\boldsymbol{s}_i} = (u_1,u_2,\ldots,u_M)$, with a $1$ in the $m$-th entry if the message $m \in \mathcal{M}_{\boldsymbol{s}_1,\ldots,\boldsymbol{s}_i}$. Given a binary sequence representation $\boldsymbol{s} = (\boldsymbol{s}_1, \boldsymbol{s}_2, \ldots, \boldsymbol{s}_C)$ of a message $m \in \mathcal{M}$, we define a set of $C$ vectors $\mathcal{U} = \{\boldsymbol{u}_{\boldsymbol{s}_1}, \boldsymbol{u}_{\boldsymbol{s}_1,\boldsymbol{s}_2}, \ldots, \boldsymbol{u}_{\boldsymbol{s}_1,\boldsymbol{s}_2,\ldots,\boldsymbol{s}_C}\}$. Finally, we can reuse \eqref{eq4} and \eqref{eq3} by inserting the appropriate $\mathcal{U}$ and $\boldsymbol{u}_{\boldsymbol{s}_1,\ldots,\boldsymbol{s}_j}$. 


\subsection{Flexible AE-Based UEP Code Design}
\label{AE_UEP_design}

We exemplify the AE-based UEP code design for $M=16$ messages transmitted using $n=7$ channel uses over the AWGN channel for $C=2$ error-protection classes.  
For both message-wise and bit-wise UEP, we design pairs $(f,g)$ that explore the trade-off between per-class error probabilities $(P_\mathrm{e}^{(1)},P_\mathrm{e}^{(2)})$. The distinguishing feature of the proposed AE-based design is that the trade-off among the values in $\boldsymbol{P}_\mathrm{e}$ is flexibly achieved by controlling the weighting factors $\lambda_1=\lambda$ and $\lambda_2=1-\lambda$, $\lambda \in [0,1]$.

\subsubsection{Message-wise UEP}

For message-wise UEP, we partition the set of $M=16$ messages into two classes, each containing half of the messages, i.e., $|\mathcal{M}_1|=8$ and $|\mathcal{M}_2|=8$. In
Fig. \ref{Fig_2} (solid lines), we plot the error-probability profile $\boldsymbol{P}_\mathrm{e} = (P_\mathrm{e}^{(1)},P_\mathrm{e}^{(2)})$ of the trained AE-based encoder-decoder pair $(f,g)$ for  $E_\mathrm{b}/N_0=1,3,5$, and $7$ dB, where $E_\mathrm{b}$ denotes the energy per bit, $N_0$ is the noise power spectral density, and $\sigma^2=(2RE_\mathrm{b}/N_0)^{-1}$. Each curve is obtained by evaluating $(P_\mathrm{e}^{(1)},P_\mathrm{e}^{(2)})$ for the weight parameter $\lambda=\{0.1,0.2,\ldots,0.9\}.$
As in \cite{OShea_2017}, we consider both encoder and decoder layers consisting of a single hidden layer with $M=16$ neurons, while the bottleneck layer has $n=7$ neurons. The neural network is trained at $E_\mathrm{b}/N_0=3$ dB, optimized using SGD with Adam optimizer. The learning rate is $\alpha = 0.001 $, $\beta_{1} = 0.9$ and $\beta_2 = 0.999$.

First, note that for $\lambda=0.5$, the values of $(P_\mathrm{e}^{(1)},P_\mathrm{e}^{(2)})$ coincide, as our AE code boils down to the equal error protection AE code presented in \cite{OShea_2017}. Secondly, note that the  curves $(P_\mathrm{e}^{(1)},P_\mathrm{e}^{(2)})$ are symmetric with respect to $\lambda=0.5$. Finally, as we increase $\lambda$ from $0.5$ toward $1$, for any fixed $E_\mathrm{b}/N_0$, the AE-based UEP code error-probability profile $\boldsymbol{P}_\mathrm{e} = (P_\mathrm{e}^{(1)},P_\mathrm{e}^{(2)})$ sweeps through a sequence of pairs where $P_\mathrm{e}^{(1)}$ gradually improves, while $P_\mathrm{e}^{(2)}$ gradually deteriorates. 
Thus, the parameter $\lambda$ offers a flexible ``fine-tuning knob'' that allows for the selection of the desired trade-off within the region $\mathcal{P}_{\mathcal{W}}(\mathcal{M},n)$.


\begin{figure}[t]
	\begin{tikzpicture}
  	\begin{semilogyaxis}[width=1\columnwidth, height=7.5cm, 
	legend style={at={(0.5,0.2)}, anchor= north,font=\scriptsize, legend style={nodes={scale=0.99, transform shape}}},
   	legend cell align={left},
	legend columns=2,   	 
   	x tick label style={/pgf/number format/.cd,
   	set thousands separator={},fixed},
   	y tick label style={/pgf/number format/.cd,fixed, precision=2, /tikz/.cd},
   	xlabel={$\lambda$},
   	ylabel={$P_{\textrm{e}}$},
   	label style={font=\footnotesize},
   	grid=major,   	
   	xmin = 0, xmax = 1,
   	ymin=0.000001, ymax=1,
   	line width=0.8pt,
   	tick label style={font=\footnotesize},]
   	\addplot[blue, mark=square] 
   	table [x={x}, y={y}] {./Fig/Fig. 3/c1_1dB.txt};
   	\addlegendentry{$E_\mathrm{b}/N_0$ = 1 dB}
   	\addplot[red, mark=x] 
   	table [x={x}, y={y}] {./Fig/Fig. 3/c1_3dB.txt};
   	\addlegendentry{$E_\mathrm{b}/N_0$ = 3 dB}
   	
   	\addplot[green, mark=square] 
   	table [x={x}, y={y}] {./Fig/Fig. 3/c1_5dB.txt};
   	\addlegendentry{$E_\mathrm{b}/N_0$ = 5 dB}
   	
   	\addplot[purple, mark=triangle] 
   	table [x={x}, y={y}] {./Fig/Fig. 3/c1_7dB.txt};
   	\addlegendentry{$E_\mathrm{b}/N_0$ = 7 dB}
   	
   	\addplot[dashed,blue, mark=square, mark options={solid}] 
   	table [x={x}, y={y}] {./Fig/Fig. 3/c2_1dB.txt};
   	
   	\addplot[dashed,red, mark=x, mark options={solid}] 
   	table [x={x}, y={y}] {./Fig/Fig. 3/c2_3dB.txt};
   	\addplot[dashed,green, mark=square, mark options={solid}] 
   	table [x={x}, y={y}] {./Fig/Fig. 3/c2_5dB.txt};
   	\addplot[dashed,purple, mark=triangle, mark options={solid}] 
   	table [x={x}, y={y}] {./Fig/Fig. 3/c2_7dB.txt};
 	\end{semilogyaxis}
	\end{tikzpicture}
	\vspace*{-8mm}
	\caption{$(P_\mathrm{e}^{(1)},P_\mathrm{e}^{(2)})$ performance ($P_\mathrm{e}^{(1)}$ solid curves, $P_\mathrm{e}^{(2)}$ dashed curves) of AE-based progressive bit-wise UEP codes ($C=2$, $k_1=2$, $k_2=2$, $n=7$) for different values of $\lambda$ and $E_b/N_0 = \{1,3,5,7\}$~dB.}
	\label{Fig_3}
\end{figure}
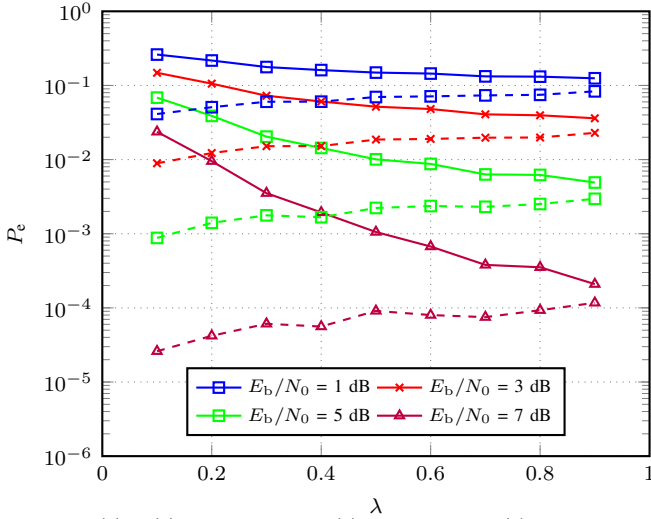 

\subsubsection{Bit-wise UEP} 

In this scenario, $M=16$ messages are represented as $k=4$-bit sequences $\boldsymbol{s} \in (\mathcal{S}_1,\mathcal{S}_2)$, where submessage $\boldsymbol{s}_1 \in \mathcal{S}_1$ contains the first two bits ($k_1=2$), while submessage $\boldsymbol{s}_2 \in \mathcal{S}_2$ contains the remaining two bits ($k_2=2$). Fig. \ref{Fig_2} (dashed lines) plots the error-probability profile $\boldsymbol{P}_\mathrm{e} = (P_\mathrm{e}^{(1)},P_\mathrm{e}^{(2)})$ of a trained AE-based encoder-decoder pair $(f,g)$ for the same values of $E_\mathrm{b}/N_0$ and for the same sequence of $\lambda$ values as in the message-wise UEP example. The encoder and decoder pairs $(f,g)$ are obtained using the same architecture and training methodology as for the message-wise UEP. 
Notably, the error-probability profile $\boldsymbol{P}_\mathrm{e}$ shows a similar behavior as for the case of message-wise UEP, despite the fact that the two scenarios are fundamentally different.

For the progressive bit-wise UEP, where the success in decoding the second message subblock is conditioned on successful decoding of the first subblock, the error-probability profile $\boldsymbol{P}_\mathrm{e}$ changes as illustrated in Fig. \ref{Fig_3}. Informally, as $\lambda \rightarrow 0$, the codewords corresponding to $\mathcal{M}_{\boldsymbol{s}_1}$ (for any $\boldsymbol{s}_1$) converge to each other, while as $\lambda \rightarrow 1$ they diverge from each other, approaching equal error protection. Note that the resulting code design is related to the problems of superposition coding for degraded broadcast channels \cite{Cover_1998} and designing UEP modulation constellations \cite{Wei_1993}.


\begin{figure}[t]
	\begin{tikzpicture}
  	\begin{loglogaxis}[width=1\columnwidth, height=7cm, 
	legend style={at={(0.71,0.98)}, anchor= north,font=\scriptsize, legend style={nodes={scale=0.99, transform shape}}},
   	legend cell align={left},
	legend columns=1,   	 
   	x tick label style={/pgf/number format/.cd,fixed,
   	 precision=1, /tikz/.cd},
   	y tick label style={/pgf/number format/.cd,fixed, precision=1, /tikz/.cd},
   	xlabel={$P_{\textrm{e}}^{(1)}$},
   	ylabel={$P_{\textrm{e}}^{(2)}$},
   	label style={font=\footnotesize},
   	grid=major,   	
   	xmin =0.00001, xmax = 1,
   	x dir=reverse,
   	ymin=0.00001, ymax=1,
   	y dir=reverse,
   	line width=0.85pt,
   	tick label style={font=\footnotesize},]
    \addplot[blue, mark=square*] 
   	table [x={x}, y={y}] {./Fig/Fig. 4/7dB.txt}; 
   	\addlegendentry{AE message-wise UEP Codes} 
   	\addplot[red, only marks, mark=*] 
   	table [x={x}, y={y}] {./Fig/Fig. 4/Scatter.txt}; 
   	\addlegendentry{Random Coset UEP Codes}
   	\node at (-1.8,-10.4) {\small $\lambda=0.1$};
   	\node at (-10,-1.5) {\small $\lambda=0.9$};
  	\end{loglogaxis}
	\end{tikzpicture}
	\vspace*{-8.5mm}
	\caption{Comparison of $(P_\mathrm{e}^{(1)},P_\mathrm{e}^{(2)})$ performance of AE-based progressive bit-wise UEP codes vs random coset UEP codes ($C=2$, $|\mathcal{M}_1|=8, |\mathcal{M}_2|=8$, $n=7$) for $E_b/N_0=7$ dB.}
	\label{Fig_new}
\end{figure}
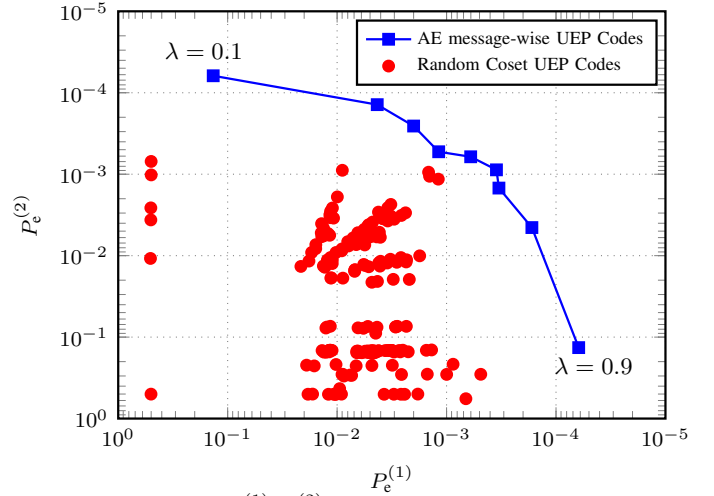

\section{Performance of AE-based UEP Codes}

In this section, we compare the AE-based UEP codes against selected classes of UEP codes.

\subsubsection{Message-wise UEP} In \cite{Shkel_2015}, the authors consider a message-wise UEP construction based on a union of coset codes. For each message class $\mathcal{M}_i$, we generate a random binary generator matrix $\boldsymbol{G}_i$ of dimension $k_i \times n$ and a random binary shift vector $\boldsymbol{v}_i$ of length $n$. For each message $m \in \mathcal{M}_i$, the corresponding codeword $\boldsymbol{x}$ is obtained as the binary phase shift keying (BPSK) modulated version of $\boldsymbol{s}\boldsymbol{G}_i+\boldsymbol{v}_i,$ where $\boldsymbol{s}$ is the binary sequence representation of $m$. For $C=2$ and $n=7$, in order to generate $\mathcal{M}_1=\mathcal{M}_2=8$ codewords, we set $k_1=2$ and $k_2=2$. We compare the AE-based design with a randomly generated set of $200$ random coset-based message-wise UEP codes. Note that, unlike the AE-based design where we use $\lambda$ to control the trade-off between per-class error probabilities in $\boldsymbol{P}_\mathrm{e}$, for random coset-based codes such a control of $\boldsymbol{P}_\mathrm{e}$ is not trivial, thus we compare against a sample of $200$ randomly generated codes. Although the coset-based design is asymptotically good for the BSC and the BEC \cite{Shkel_2015}, for the AWGN and short code lengths, the performance of the best selected candidates does not match that of AE-based designed codes (note that coset-based codes use the encoding function $f: \mathcal{M} \rightarrow \{+1,-1\}^n$). Fig. \ref{Fig_new} demonstrates that, for $\lambda = \{0.1, 0.2, \ldots, 0.8, 0.9\}$, the error-probability pairs $(P_\mathrm{e}^{(1)},P_\mathrm{e}^{(2)})$ for AE-based codes consistently outperform randomly-sampled set containing $200$ coset-based codes (at $E_\mathrm{b}/N_0=7$ dB). In other words, AE-based codes perform consistently closer to the boundary of the region $\mathcal{P}_{\mathcal{W}}(\{\mathcal{M}_i\}_{i=1}^C,n)$. 

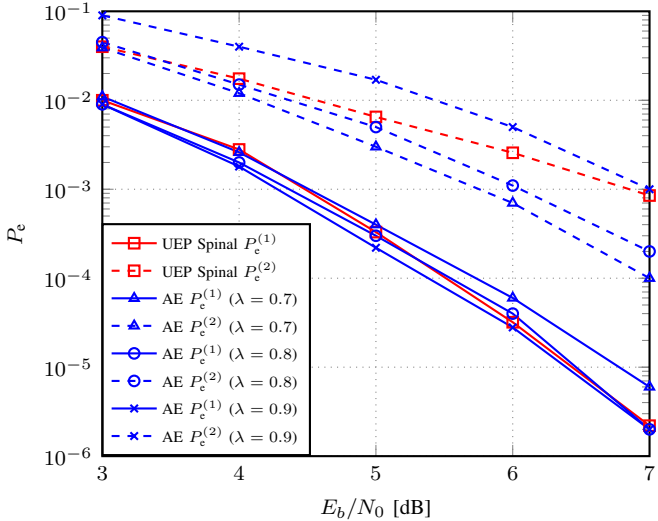
\begin{figure}[t]
	\begin{tikzpicture}
  	\begin{semilogyaxis}[width=1\columnwidth, height=7.5cm, 
	legend style={at={(0.19,0)}, anchor= south,font=\scriptsize, legend style={nodes={scale=0.85, transform shape}}},
   	legend cell align={left},
	legend columns=1,   	 
   	x tick label style={/pgf/number format/.cd,
   	set thousands separator={},fixed},
   	y tick label style={/pgf/number format/.cd,fixed, precision=2, /tikz/.cd},
   	xlabel={$E_b/N_0$ [dB]},
   	ylabel={$P_{\textrm{e}}$},
   	label style={font=\footnotesize},
   	grid=major,   	
   	xmin = 3, xmax = 7,
   	ymin=0.000001, ymax=0.1,
   	line width=0.8pt,
   	tick label style={font=\footnotesize},]
    \addplot[red, mark=square] 
   	table [x={x}, y={y}] {./Fig/Fig. 6/ca1_conv.txt};
   	\addlegendentry{UEP Spinal $P_{\textrm{e}}^{(1)}$ }
   	\addplot[dashed,red, mark=square, mark options={solid}] 
   	table [x={x}, y={y}] {./Fig/Fig. 6/ca2_conv.txt};
   	\addlegendentry{UEP Spinal $P_{\textrm{e}}^{(2)}$ }
   	\addplot[blue, mark=triangle] 
   	table [x={x}, y={y}] {./Fig/Fig. 6/cl1_prog05_03.txt};
   	\addlegendentry{AE $P_{\textrm{e}}^{(1)}$ ($\lambda=0.7$)}
   	\addplot[dashed,blue, mark=triangle, mark options={solid}] 
   	table [x={x}, y={y}] {./Fig/Fig. 6/cl2_prog05_03.txt};
   	\addlegendentry{AE $P_{\textrm{e}}^{(2)}$ ($\lambda=0.7$)}
   	\addplot[blue, mark=o] 
   	table [x={x}, y={y}] {./Fig/Fig. 6/cl1_prog5_02.txt};
   	\addlegendentry{AE $P_{\textrm{e}}^{(1)}$ ($\lambda=0.8$)}
   	\addplot[dashed, blue, mark=o, mark options={solid}] 
   	table [x={x}, y={y}] {./Fig/Fig. 6/cl2_prog5_02.txt};
   	\addlegendentry{AE $P_{\textrm{e}}^{(2)}$ ($\lambda=0.8$)}
   	\addplot[blue, mark=x] 
   	table [x={x}, y={y}] {./Fig/Fig. 6/cl1_prog_05_01.txt};
   	\addlegendentry{AE $P_{\textrm{e}}^{(1)}$ ($\lambda=0.9$)}
   	\addplot[dashed,blue, mark=x, mark options={solid}] 
   	table [x={x}, y={y}] {./Fig/Fig. 6/cl2_prog05_01.txt};
   	\addlegendentry{AE $P_{\textrm{e}}^{(2)}$ ($\lambda=0.9$)}
 	\end{semilogyaxis}
	\end{tikzpicture}
	\vspace*{-7.25 mm}
	\caption{Comparison of $(P_\mathrm{e}^{(1)},P_\mathrm{e}^{(2)})$ vs $E_b/N_0$ performance of AE-based and spinal bit-wise UEP codes ($C=2$, $k_1=4$, $k_2=10$, $n=32$).}
	\label{Fig_5}
\end{figure}


 \subsubsection{Bit-wise UEP} We compare the AE-based bit-wise UEP codes with the bit-wise UEP spinal codes \cite{Yu_2016}. The code parameters are $C=2$, $k_1=4$, $k_2=10$, and  $n=32$. For the AE-based code, we apply the AE architecture with a single hidden layer with 500 neurons for both the transmitter and the receiver and $n=32$ neurons for the bottleneck layer. Furthermore, we consider $\lambda = \{0.7,0.8,0.9\}$.  The training procedure is the same as described in Section~\ref{AE_UEP_design} except that the training is done at $E_\mathrm{b}/N_0=1$ dB.   The AE-based codes perform comparably to spinal codes in terms of $P_{\textrm{e}}^{(1)}$, while significantly outperforming them for $P_{\textrm{e}}^{(2)}$.

\begin{figure}[t]
	\begin{tikzpicture}
  	\begin{loglogaxis}[width=0.95\columnwidth, height=7.5cm, 
	legend style={at={(0.3,0.4)}, anchor=north,font=\scriptsize},
   	legend cell align={left},
	legend columns=1,   	 
   	x tick label style={/pgf/number format/.cd,fixed,
   	 precision=1, /tikz/.cd},
   	y tick label style={/pgf/number format/.cd,fixed, precision=1, /tikz/.cd},
   	xlabel={$P_{\textrm{e}}^{(1)}$},
   	ylabel={$P_{\textrm{e}}^{(2)}$},
   	label style={font=\footnotesize},
   	grid=major,   	
   	xmin =0.00005, xmax = 1,
   	x dir=reverse,
   	ymin=0.0009, ymax=1,
   	y dir=reverse,
   	line width=0.8pt,
   	tick label style={font=\footnotesize},]
    \addplot[green, mark=square] 
   	table [x={y}, y={x}] {./Fig/Fig. 6_1/error_7_4.txt}; 
   	\addlegendentry{AE-based code (4,7)} 
   	\addplot[dashed,green, mark=square, mark options={solid}] 
   	table [x={x}, y={y}] {./Fig/Fig. 6_1/Linear_7_4.txt}; 
   	\addlegendentry{Superposition code (4,7)} 
   	\addplot[blue, mark=o] 
   	table [x={y}, y={x}] {./Fig/Fig. 6_1/error_14_8.txt}; 
   	\addlegendentry{AE-based code (8,14)} 
   	\addplot[dashed,blue, mark=o, mark options={solid}] 
   	table [x={x}, y={y}] {./Fig/Fig. 6_1/Linear_14_8.txt}; 
   	\addlegendentry{Superposition code (8,14)} 
   	\addplot[red, mark=x] 
   	table [x={y}, y={x}] {./Fig/Fig. 6_1/error_21_12.txt}; 
   	\addlegendentry{AE-based code (12,21)} 
   	\addplot[dashed,red, mark=x, mark options={solid}] 
   	table [x={x}, y={y}] {./Fig/Fig. 6_1/Linear_21_12.txt};
   	\addlegendentry{Superposition code (12,21)}
   	\node at (-0.88,-6.4) {\small $\lambda=0.1$};
   	\node at (-9,-3.1) {\small $\lambda=0.9$};
  	\end{loglogaxis}
	\end{tikzpicture}
	\vspace*{-3.9mm}
	\caption{Comparison of $(P_\mathrm{e}^{(1)},P_\mathrm{e}^{(2)})$ performance of AE-based and superposition of random Gaussian codes for $(k,n)$-pairs: $(4,7)$, $(8,14)$ and $(12,21)$, where $k_1=\frac{1}{4}k$ and $k_2=\frac{3}{4}k$ at $E_b/N_0=5$~dB.}
	\label{Fig_4}
\end{figure}
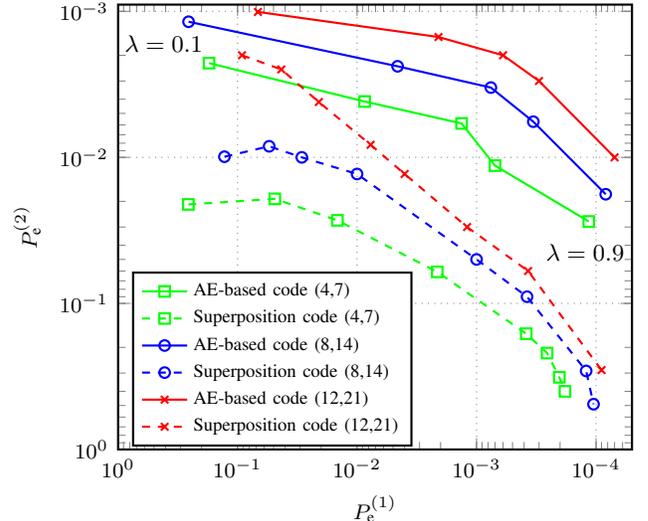

Next, we compare the AE-based design with the bit-wise UEP codes based on the superposition of random Gaussian codes \cite{Karimzadeh_2019}.\footnote{We consider a special case of \cite{Karimzadeh_2019}, where decoding of both message classes is attempted after all $n$ codeword symbols are received.} For $C=2$, each codeword from a set of $2^{k_1}$ random Gaussian codewords whose symbols are drawn from $\mathcal{N}(0,\sigma^2_1)$ is superimposed by a set of $2^{k_2}$ random Gaussian codewords whose symbols are sampled from $\mathcal{N}(0,\sigma^2_2)$. The resulting set of $M=2^k$ codewords is normalized to obey the total energy constraint. We consider three equal-rate $(k,n)$-pairs, $(4,7)$, $(8,14)$, and $(12,21)$, where $k_1=\frac{1}{4}k$ and $k_2=\frac{3}{4}k$. We use a single parameter $\mu$ to design the superposition of random Gaussian codes by varying $(\sigma^2_1,\sigma^2_2) =(\mu,1-\mu)$, providing a similar control of the trade-off between error probabilities $(P_\mathrm{e}^{(1)},P_\mathrm{e}^{(2)})$ as $\lambda$ provides for the AE-based codes. Fig. \ref{Fig_4} demonstrates superior performance of the AE-based codes for $\lambda=\{0.1,0.3,0.5,0.7,0.9\}$ at $E_\mathrm{b}/N_0=5$ dB compared to the best performing UEP codes based on the superposition of random Gaussian codes for $\mu=\{0.3,0.4,0.5,0.6,0.7\}$ (from a set of $200$ randomly generated codes).

\section{Conclusion}

We presented a novel learning-based approach to the design of UEP codes using deep AEs. The design is based on a generalized AE loss function that accommodates both message-wise and bit-wise UEP code design. The proposed AE-based codes show superior performance to known UEP approaches, such as random coset codes or superposition of random Gaussian codes. They also outperform state-of-the-art UEP rateless spinal codes. Besides excellent performance, AE-based UEP codes provide a built-in flexible mechanism for weighting loss function components that results in a graceful trade-off between per-class error probabilities.

\begin{thebibliography}{1}

\bibitem{OShea_2017}
T. O’Shea and J. Hoydis, ``An introduction to deep learning for the physical layer,'' \emph{IEEE Trans. Cogn. Commun. Netw.,} vol. 3, no. 4, pp. 563-575, Dec. 2017.

\bibitem{OShea_2017_2}
T. O'Shea, T. Erpek, and T.C. Clancy, ``Deep learning based MIMO communications,'' Jul. 2017., arXiv:1707.07980v1 [cs.IT] . [Online]. Available: https://arxiv.org/abs/1707.07980

\bibitem{Dorner_2018}
S. Dörner, S. Cammerer, J. Hoydis, and S. ten Brink, ``Deep learning based communication over the air,'' \emph{IEEE J. Sel. Topics Signal Process.,}  vol. 12, no. 1, pp. 132-143, Feb. 2018. 

\bibitem{Masnick_1967}
B. Masnick and J.  Wolf, ``On linear unequal error protection codes,'' \emph{IEEE Trans. Inf. Theory,} vol. 13, no. 4, pp. 600-607, Oct. 1967.

\bibitem{Shkel_2015}
Y.Y. Shkel, V.Y. Tan, and S.C. Draper, ``Unequal message protection: Asymptotic and non-asymptotic tradeoffs,'' \emph{IEEE Trans. Inf. Theory,} vol. 61, no. 10, pp. 5396-5416, Oct. 2015.

\bibitem{Borade_2009}
S. Borade, B. Nakiboglu, and L. Zheng, ``Unequal error protection: An information-theoretic perspective,'' \emph{IEEE Trans. Inf. Theory,} vol. 55, no. 12, pp. 5511-5539, Dec. 2009.

\bibitem{Pishro-Nik_2005}
H. Pishro-Nik, N. Rahnavard, and F. Fekri, ``Nonuniform error correction using low-density parity-check codes,'' \emph{IEEE Trans. Inf. Theory,}  vol. 51, no. 7, pp. 2702-2714, July 2005.

\bibitem{Sejdinovic_2009}
D. Sejdinovic, D. Vukobratovic, A. Doufexi, V. Senk, and R.J. Piechocki, ``Expanding window fountain codes for unequal error protection,'' \emph{IEEE Trans. Commun.,} vol. 57, no. 9, pp. 2510-2516, Sep. 2009.

\bibitem{Yu_2016}
X. Yu, Y. Li, W. Yang, and Y. Sun, ``Design and analysis of unequal error protection rateless spinal codes,'' \emph{IEEE Trans. Commun.,} vol. 64, no. 11, pp. 4461-4473, Nov. 2016.

\bibitem{Karimzadeh_2019}
M. Karimzadeh and M. Vu, ``Short Blocklength Priority-Based Coding for Unequal Error Protection in the AWGN Channel,'' in \emph{Proc. 2019 IEEE Global Commun. Conf. (GLOBECOM),} Waikoloa, HI, USA, Dec. 9-13, 2019, pp. 1-6.

\bibitem{adam}
D. P. Kingma and J. L. Ba, ``Adam: A method for stochastic optimization,'' in \emph{Proc. Int. Conf. on Learn. Representation,} San Diego, CA, USA, May 7-9, pp. 1-41, 2015.


\bibitem{Chande_2000}
V. Chande and N. Farvardin, ``Progressive transmission of images over memoryless noisy channels,'' \emph{IEEE J. Sel. Areas Commun.
,} vol. 18, no. 6, pp. 850--860, June 2000.

\bibitem{Cover_1998}
T.M. Cover, ``Comments on broadcast channels,'' \emph{IEEE Trans. Inf. Theory,} vol. 44, no. 6, pp. 2524-2530, Oct. 1998.

\bibitem{Wei_1993}
L.F. Wei, ``Coded modulation with unequal error protection,'' \emph{IEEE Trans. Commun.,} vol. 41, no. 10, pp. 1439-1449, Oct. 1993.

\end{thebibliography}
\end{document}